\documentclass{svjour3}                      
\smartqed  
\usepackage{graphicx}
\usepackage{amsmath}
 \journalname{Accepted in Journal of Optics (2013) of}

\pdfpagesattr{/CropBox [10 150 476 815]}

\begin{document}

\title{Twin image elimination in digital holography by combination of 
Fourier transformations
}

\titlerunning{Twin image elimination in digital holography}        

\author{Debesh Choudhury         \and
        Gautam Lohar 
}


\institute{Debesh Choudhury \at
               Department of Electronics and Communication Engineering \\
               Neotia Institute of Technology, Management and Science\\
               PO - Amira, D. H. Road, South 24 Parganas, Pin 743368, 
               West Bengal, India\\
              \email{debesh[AT]iitbombay[DOT]org}           
           \and
           Gautam Lohar \at
              Department of Electronics and Communication Engineering,
              JIS College of Engineering\\ Block A, Phase III,
              Kalyani, Nadia, Pin 741235, West Bengal, India. 
}

\date{}

\maketitle

 \begin{abstract}
 We present a new technique for removing twin image in in-line digital 
Fourier holography using a combination of Fourier transformations. 
Instead of recording only a Fourier transform hologram of the object, we 
propose to record a combined Fourier transform hologram by 
simultaneously recording the hologram of the Fourier transform and the 
inverse Fourier transform of the object with suitable weighting 
coefficients. Twin image is eliminated by appropriate inverse combined 
Fourier transformation and proper choice of the weighting coefficients. 
An optical configuration is presented for recording combined Fourier 
transform holograms. Simulations demonstrate the feasibility of twin 
image elimination. The hologram reconstruction is sensitive to phase 
aberrations of the object, thereby opening a way for holographic phase 
sensing.

 \vglue 12pt
 \end{abstract}

\section{Introduction}
 Twin image is an age old problem in holography since its invention by 
Gabor~\cite{gabor}. In 1951, Bragg and Rogers first eliminated the 
unwanted twin image by recording two holograms and doubling the 
object-to-hologram distance in between the 
exposures~\cite{rogers:nature51}. A nice review details several 
available techniques for getting rid of the twin image~\cite{twin:rev}. 
The off-axis holography of Leith and Upatnieks is the simplest method, 
but it requires high resolution recording 
materials~\cite{leith_upatnieks}. Most other methods~\cite{twin:rev} 
either used optical/digital spatial filtering, or needed to record 
multiple phase shifted holograms, or utilized iterative reconstruction 
of the holograms thereby making the recording and/or reconstruction 
process slow. We propose to surmount this problem by recording a 
combined Fourier transform hologram and its computer reconstruction by 
inverse combined Fourier transformation.

\section{Combination of Fourier transformations}
 If $g(x,y)$ is an object function, its forward Fourier transform (FT), 
i.e., ${\mathcal FT} \{g(x,y)\}$, is given by
 \begin{eqnarray}
 \int\limits_{-\infty}^{+\infty} 
\int\limits_{-\infty}^{+\infty} g(x,y) \exp \{-i2\pi (ux + vy)\} dx dy = 
G(u,v) \label{ft}
 \end{eqnarray}

\noindent and its inverse FT, i.e., ${\mathcal IFT} \{g(x,y)\}$, is 
given by
 \begin{eqnarray}
 \int\limits_{-\infty}^{+\infty} 
\int\limits_{-\infty}^{+\infty} g(x,y) \exp \{i2\pi (ux + vy)\} dx dy = 
G^\star(u,v) \label{ift}
 \end{eqnarray}

\noindent where $x,y$ and $u,v$ are the space and spatial frequency 
coordinates pairs, $i=\sqrt{-1}$, $\mathcal FT$, $\mathcal IFT$ signify 
Fourier transform and imverse Fourier transform operator and the symbol 
$\star$ stands for complex conjugation. Following the definitions of FT 
and inverse FT, the identities below also hold
 \begin{eqnarray}
 {\mathcal IFT}\{g(x,y)\} = G^{\star}(u,v) = G(-u,-v) \label{id1}\\
 {\mathcal IFT}\{g(-x,-y)\} = G^\star(-u,-v) =G(u,v) \label{id2}
 \end{eqnarray}

\noindent   We now define a combined Fourier transform (CFT) 
as~\cite{cft:ansari}
 \begin{eqnarray} G_{cft}(u,v) & =& a_1 {\mathcal FT}[g(x,y)] + a_2 
{\mathcal IFT}[g(x,y)] \nonumber \\ & = & a_1 G(u,v) + a_2 G(-u,-v) 
\label{cft}
 \end{eqnarray}

\noindent where $a_1$ and $a_2$ are two constant coefficients, real or 
complex, but must satisfy $a_1^2 \neq a_2^2$. 
The object function $g(x,y)$ can be recovered by an inverse CFT (ICFT) 
given by
 \begin{eqnarray}
 g(x,y) & =& (a_1^2 - a_2^2)^{-1} \large[ a_1 
{\mathcal IFT}\{G_{cft}(u,v)\} \nonumber \\ && - \ a_2 {\mathcal 
FT}\{G_{cft}(u,v)\} \large] \label{icft}
 \end{eqnarray}

\section{Combined FT Hologram recording}

We record a CFT hologram by adding a coherent plane wave to the wave 
field distribution of equation~(\ref{cft}). If $R$ is the amplitude of 
the collimated reference wave, the wave field distribution at the 
hologram recording plane will be given by
 \begin{equation}
 \Psi (u,v) = R + G_{cft}(u,v) \ \ \ .
 \end{equation}

\noindent The recorded hologram intensity will be given by
 \begin{eqnarray}
 I(u,v) & =& \vert \Psi(u,v) \vert^2 \nonumber \\
 & = & \vert R \vert^2 + \vert G_{cft} \vert^2 + RG_{cft}^\star + 
R^\star G_{cft} \label{holo}
 \end{eqnarray}

\noindent The amplitude of the reference wave $R$ is made large enough 
so as to make the hologram recording linear. We also capture a separate 
record of the reference wave intensity $\vert R \vert^2=R^2$ only, prior 
to recording the combined Fourier transform hologram. Subtracting $R^2$ 
from both sides of equation~(\ref{holo}) and also dividing both sides by 
$R^2$, we get the modified hologram intensity as
 \begin{eqnarray}
 I^\prime = {\vert G_{cft} \vert^2 \over R^2} + {1 \over R^2} \{ 
RG_{cft}^\star + R^\star G_{cft} \} \label{h_irr}
 \end{eqnarray}   

\noindent Since, $\vert R \vert^2 \gg \vert G_{cft} \vert^2$, $\vert 
G_{cft} \vert^2/\vert R \vert^2$ is vanishingly small, hence it can be 
neglected and equation~(\ref{h_irr}) reads
 \begin{eqnarray}
 I^\prime \approx {1 \over \vert R \vert^2} \{ RG_{cft}^\star + R^\star 
G_{cft} \} \label{h_mirr}
 \end{eqnarray}  

\section{Object reconstruction from the combined FT hologram}

The object function $g(x,y)$ can be reconstructed from the CFT hologram 
by illuminating the hologram by the reference wave $R$ and by ICFT 
operation, i.e.,
 \begin{eqnarray}
 {\mathcal ICFT} [RI^\prime] = {\mathcal ICFT} [G_{cft}^\star] + {\mathcal 
ICFT} [G_{cft}]   \label{recons}
 \end{eqnarray}

\noindent since ${R^\star / R}= 1$ because ${R^\star = R}$, and 
${\mathcal ICFT}$ stands for ICFT operator. Using the identities of 
equations~(\ref{id1}) and (\ref{id2}) and expanding the inverse CFTs, 
equation~(\ref{recons}) can be expressed as
 \begin{eqnarray}
 {\mathcal ICFT} [RI^\prime] = A_1 g(x,y) + A_2 g(-x,-y)  \label{recons1}
 \end{eqnarray}

\noindent where $A_1$ and $A_2$ are constants involving the weighting 
coefficients $a_1$ and $a_2$ given by
 \begin{eqnarray}
 A_1 = \left( 1+ {{a_1 a_2^\star - a_1^\star a_2} \over {\vert a_1 
\vert^2 - \vert a_2 \vert^2} }
 \right) \ , \
 A_2 = \left( { {a_1 a_1^\star - a_2 a_2^\star} \over {\vert a_1 \vert^2 
- \vert a_2 \vert^2} } \right)
 \end{eqnarray}

\noindent If we take values of $a_1$ and $a_2$ such that they are 
mutually complex conjugate, as for example, if 
 \begin{eqnarray}
 a_1=b_1 + i b_2 \ \ \ {\rm and} \ \ \ 
 a_2=b_1 - i b_2  \label{a1:a2}
 \end{eqnarray}

\noindent $b_1$ and $b_2$ being real, $A_1=1$ but $A_2=0$, and 
equation~(\ref{recons1}) reads
 \begin{eqnarray}
 \left\{ {\mathcal ICFT} [RI^\prime] \right\}_{a_1=a_2^\star} = g(x,y)
 \end{eqnarray}
 
\noindent That means, only the object function $g(x,y)$ is reconstructed 
and the conjugate reconstruction $g(-x,-y)$ is eliminated. If $a_1$ and 
$a_2$ are not exactly complex conjugates of each other, the twin image 
will be present. The strengths of the two images will depend on the 
real and imaginary parts of $a_1$ and $a_2$.

\noindent By De Moivre's theorem, equation~(\ref{a1:a2}) can be 
expressed as
 \begin{eqnarray}
 a_1 = B \exp(i\phi) \ \ \ {\rm and} \ \ \ 
 a_2 = B \exp(-i\phi) \label{a1a2:aphi} \\
 {\rm where} \ \  B = \sqrt{b_1^2+b_2^2} \ \ {\rm and} \ \ 
 \phi = \tan^{-1}(b_2/b_1) \label{B}
 \end{eqnarray}

\noindent Putting $a_1$ and $a_2$, equation~(\ref{cft}) becomes
 \begin{eqnarray} G_{cft}(u,v) = B \{\exp(i\phi) G(u,v) 
\nonumber \\ \ \ \ + \exp(-i\phi) G(-u,-v)\} \nonumber \\ = 
B \exp(-i\phi) \{\exp(i2\phi) G(u,v) + G(-u,-v)\} 
\label{cft:phi}
 \end{eqnarray}

\section{Optical configuration for combied FT hologram recording}

\noindent Now, we can create a wave field distribution that is 
equivalent to equation~(\ref{cft:phi}) using an optical arrangement of
 \begin{figure}[htb]
 \centerline{\includegraphics[width=12cm]{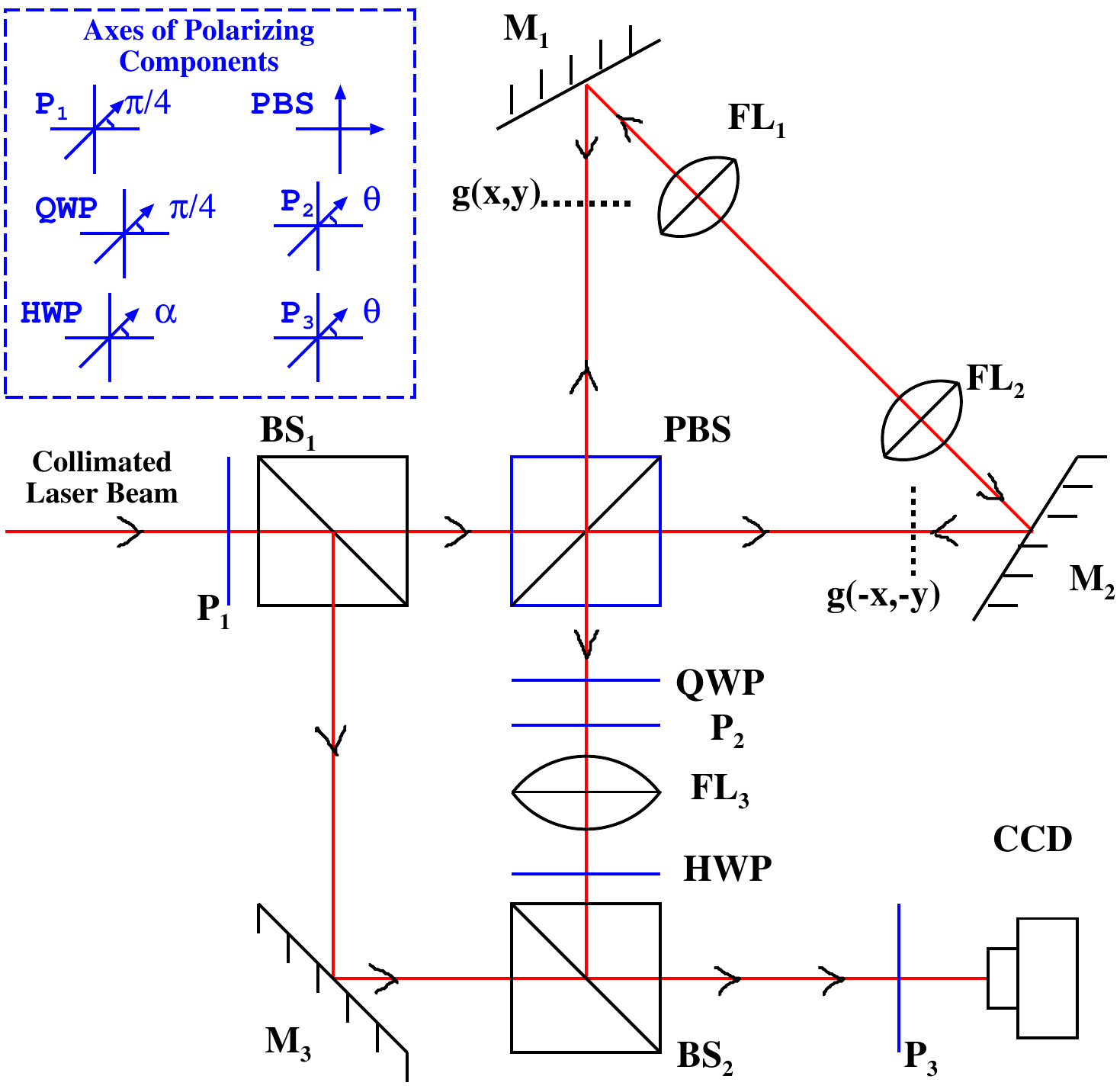}}
 \caption{Recording geometry for combined FT holograms. The polarizing 
components and their orientations are shown in the top left corner in a 
dashed box.}
 \label{interset}
 \end{figure}
 Fig.\ref{interset} and record optical CFT holograms. An expanded 
collimated beam of light from a He-Ne laser (red colored online) is 
split through a beam-splitter BS$_1$. One part is directed to a 
polarizing beam-splitter PBS. The two mirrors M$_1$ and M$_2$ form a 
triangular path interferometer, and the transmitted/reflected beams 
recombine by PBS. The transmissive object with amplitude transmittance 
$g(x,y)$ is placed inside the interferometer. We place one Fourier lens 
pair FL$_1$ and FL$_2$ of equal focal lengths inside the interferometer 
such that an inverted image of the object is formed as shown in 
Fig.\ref{interset}. The interferometric arrangement is so adjusted such 
that the object $g(x,y)$ and its inverted image $g(-x,-y)$ are 
equidistant from the beam-splitter PBS. So, one can get at the output 
side of the interferometer Fresnel propagated wave fields from (i)~the 
object function $g(x,y)$ and from (ii) an inverted version of the object 
function, i.e., $g(-x,-y)$. A third Fourier lens FL$_3$ is also placed 
which produces the Fourier transforms of $g(x,y)$ and $g(-x,-y)$ with 
appropriate coefficients that depend on the phase relationship between 
the wave fields propagating through the interferometric arms. The 
orientations of the polarizing components are shown in a dashed box. 
Another part of the input beam from the laser, which is transmitted from 
the beam-splitter BS$_1$ and is reflected by the mirror M$_3$, 
superposes with object bearing beams after transmitting through another 
beam-splitter BS$_2$. To get a CFT hologram at the CCD plane, the phase 
differences between the beams are adjusted to $2\phi$ using 
polarization-induced phase~\cite{deb:polaphase}.

The input laser beam is polarized by the polarizer P$_1$ at angle 
$\pi/4$ to the $x$-axis, and may be represented by a Jones vector $\hat 
E$ as~\cite{jones:calculus}
 \begin{eqnarray}
 \hat E = E_0 
 \begin{bmatrix} 1 \\ 1 \end{bmatrix}
 \end{eqnarray}

\noindent $E_0$ being the amplitude. This polarized beam is split by the 
polarizing beam-splitter PBS into two orthognally polarized beam along 
the $x$-axis and $y$-axis, and the Jones matrices corresponding to the 
PBS along the $x$-axis and $y$-axis may be given by~\cite{jones:calculus}
 \begin{eqnarray}
 P_x = 
 \begin{bmatrix} 1 & 0 \\ 0 & 0 \end{bmatrix}  \ , \ \ \ \ \ 
 P_y =
 \begin{bmatrix} 0 & 0 \\ 0 & 1 \end{bmatrix}
 \end{eqnarray} 

\noindent The reflected polarized beam, after a round trip through the 
cyclic interferometer gets reflected by the PBS and comes out at the 
output side. Similarly, the transmitted polarized component comes out of 
the PBS by transmission at the output side. It is to be noted that these 
two polarized beams carry the object transmittance information $g(x,y)$ 
and its inversion $g(-x,-y)$ by appropriate lens transformation inside 
the interferometer. These two orthogonally polarized image bearing beams 
face a quarter-wave retardation plate QWP whose slow axis makes an angle 
$\pi/4$ with the $x$-axis. Finally, these two polarized beams after 
passing through the quarter-wave plate will pass through an analyzer 
P$_2$ and the wave fields proceed towards the CCD camera. After Fourier 
transformation by the third lens, the vector wave field distribution at 
the CCD camera plane will be given by
 \begin{eqnarray}
 \hat G_{PI}(u,v) = P(\theta)
 [ C(\pi/4) P_x \hat E G(u,v) \nonumber \\ + \ C(\pi/4) P_y \hat E G(-u,-v) ]  
 \label{jones:eq}
 \end{eqnarray}

\noindent where $P(\theta)$ represents the Jones matrix of the analyzer 
P$_2$, $C(\pi/4)$ represents the Jones matrix of the quarter-wave 
retardation plate, and are given by~\cite{jones:calculus}
 \begin{eqnarray}
 P(\theta) &=&
 \begin{bmatrix} \cos^2\theta & \cos\theta\sin\theta \\ 
 \sin\theta\cos\theta & \sin^2\theta 
 \end{bmatrix} \\
 C(\pi/4) &=& {1\over2} \begin{bmatrix} 1+i \ \ \ & \ \ \ 1-i \\ 1-i \ 
\ \ & \ \ \ 1+i 
\end{bmatrix}
 \end{eqnarray}

\noindent Performing matrix multiplications, equation~(\ref{jones:eq}) 
can be expressed as
 \begin{eqnarray}
 \hat G_{PI}(u,v) 
 = \left[\exp\left\{i\left({\pi\over4}-\theta\right)\right\} G(u,v)
 + \exp\left\{-i\left({\pi\over4}-\theta\right)\right\}G(-u,-v)\right]  
 \nonumber \\ \times 
 \left[ \begin{array}{cc} 
 \cos\theta \\ \sin\theta \end{array} \right] \label{pi}
 \end{eqnarray}

\noindent where we have dropped off the constant amplitude factors. If 
we put $\theta=\pi/4-\phi$, the polarization-induced phase difference 
becomes $2\phi$, and equation~(\ref{pi}) becomes equivalent to the field 
distribution given by equation~(\ref{cft:phi}) and is nothing but the 
CFT of the object function $g(x,y)$ except that it is linearly 
polarized. We can also identify the phase factors 
$\exp\{i(\pi/4-\theta)\}$ and $\exp\{-i(\pi/4-\theta)\}$ as the complex 
coefficients $a_1$ and $a_2$ respectively.

Finally, an analyzer P$_3$ should be placed before the CCD camera whose 
transmission axis makes an angle $\theta$ (same as P$_2$) with the 
$x$-axis. A half-wave plate HWP should also be placed after the Fourier lens 
FL$_3$ with its slow axis making an angle $\alpha$ ($\alpha \neq 
\theta$) with the $x$-axis, so as to keeping the amplitudes of the 
object waves reaching the CCD camera much smaller than the amplitude of 
the reference wave. This will help to ensure that $\vert R \vert^2 \gg 
\vert G_{cft} \vert^2$. The object waves and the reference wave will 
transmit through the analyzer P$_3$ and interfere to form thd CFT 
hologram which can be recorded by the CCD camera.

\section{Feasibility Simulations}

\begin{figure}[htb] \hskip 1.5pt
 \centerline{\includegraphics[width=5.3cm]{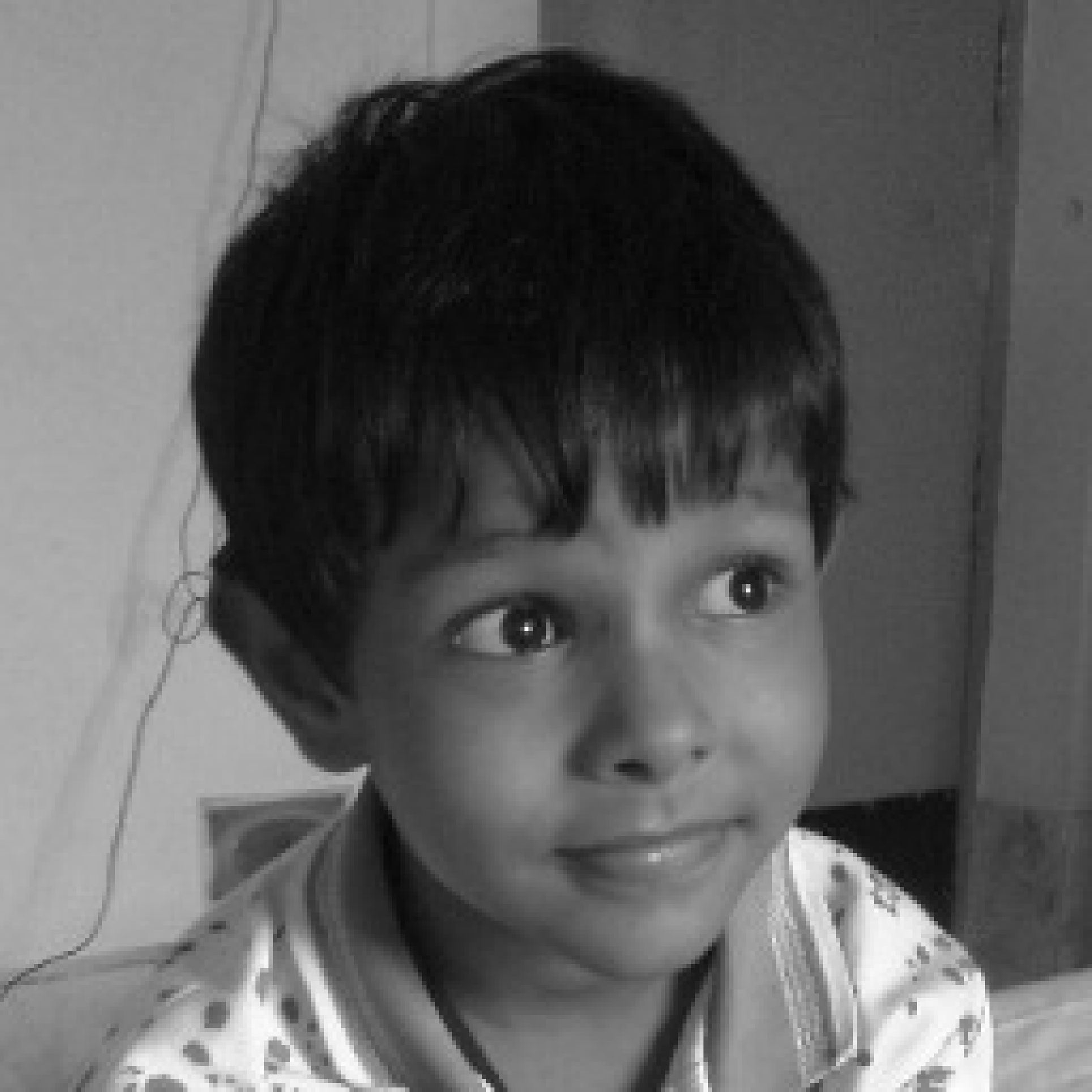} \hskip 24pt 
 \includegraphics[width=5.5cm]{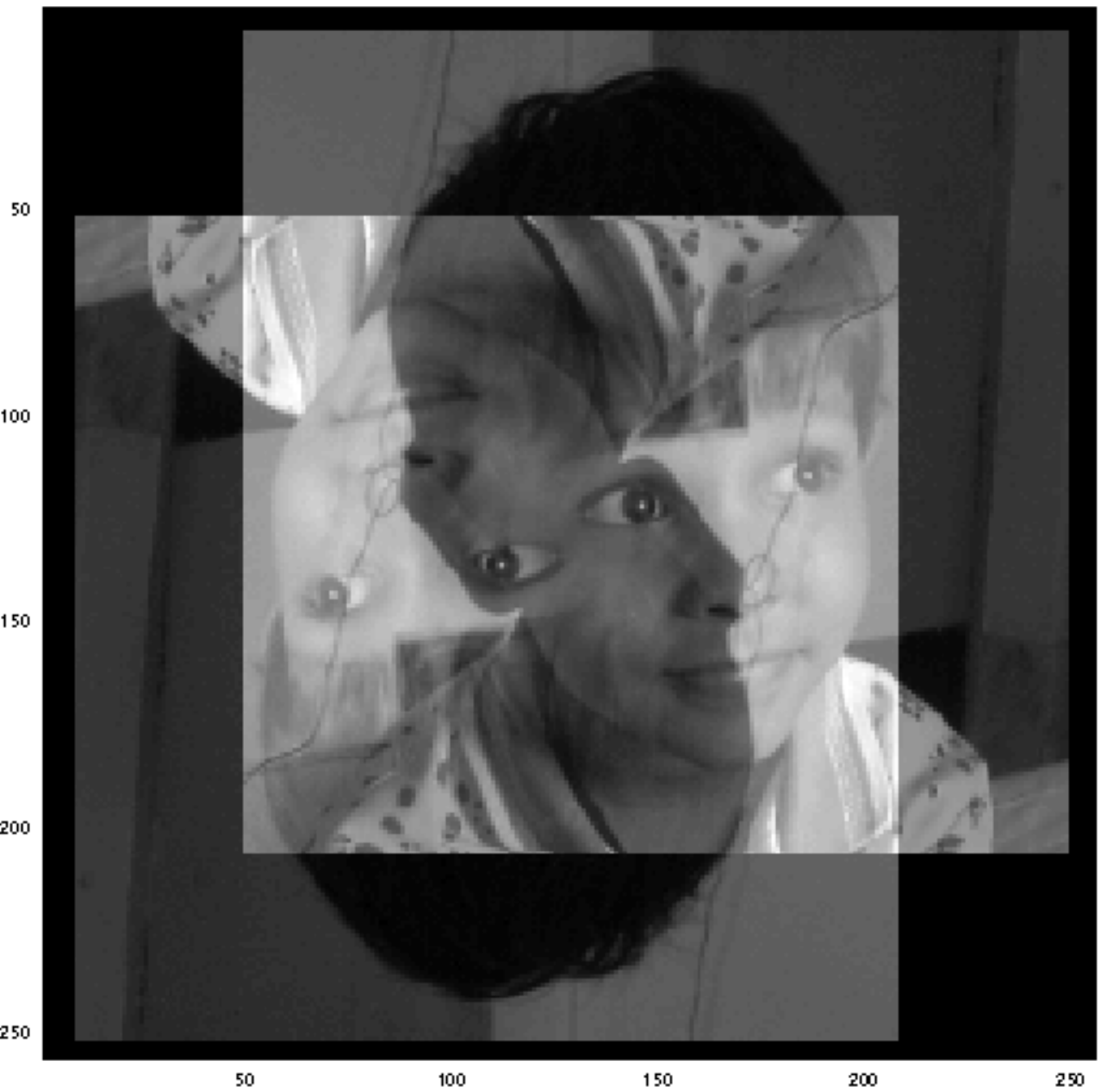} } \vskip 2pt
 \centerline{(a) \hskip 5.8cm (b)} \vskip 6pt
 \centerline{\includegraphics[width=5.5cm]{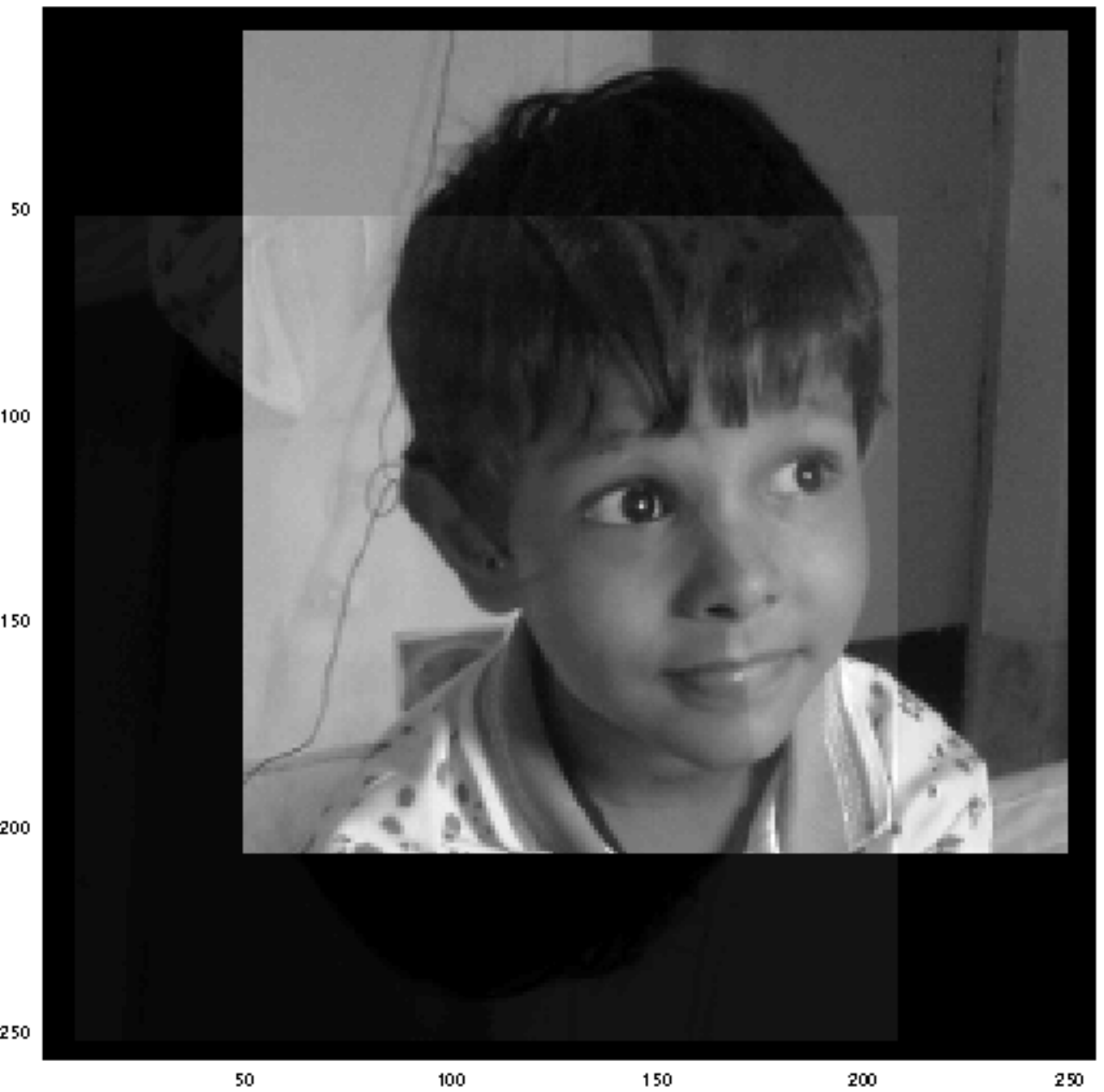} \hskip 24pt
 \includegraphics[width=5.5cm]{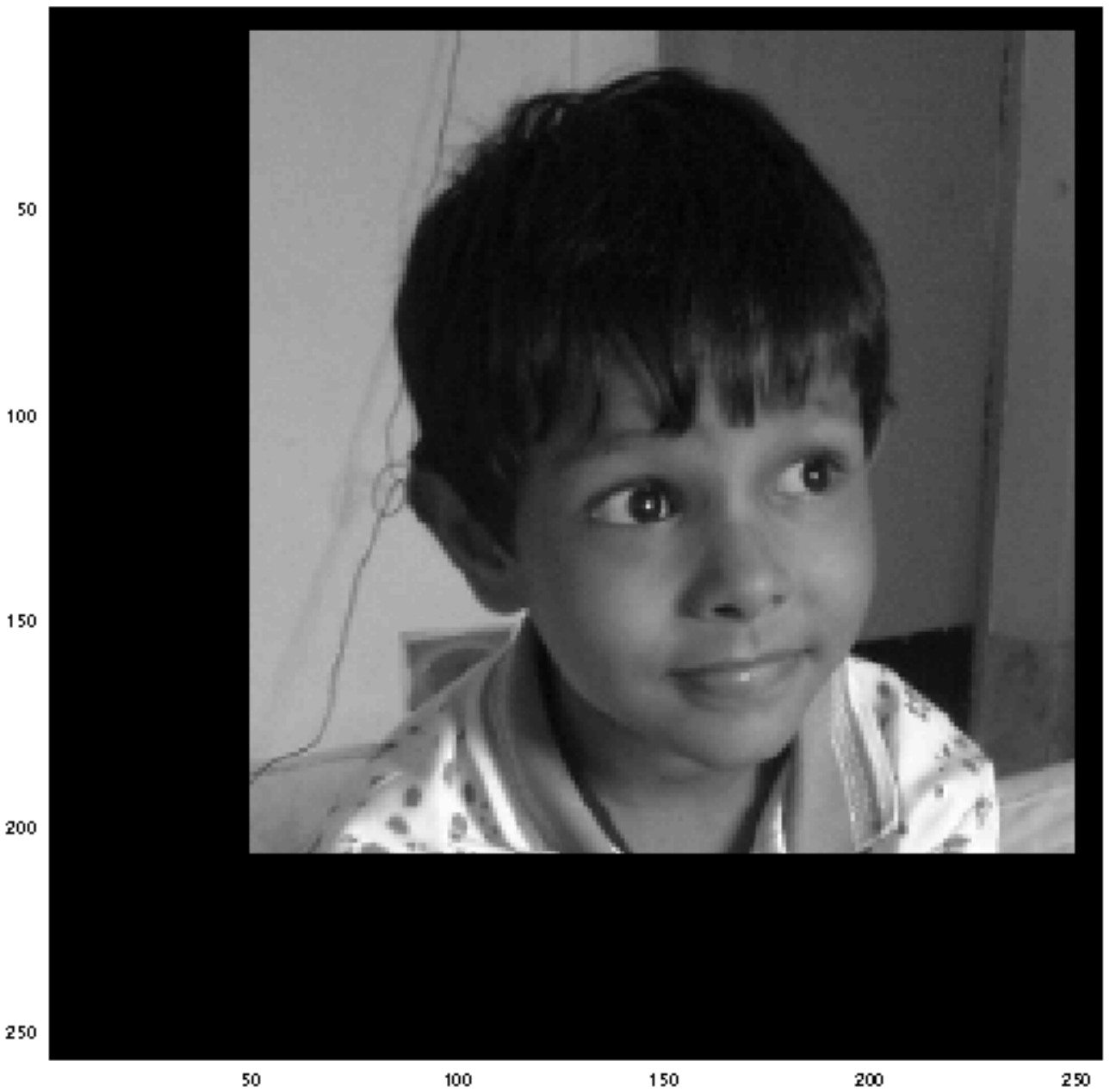} } \vskip 2pt
 \centerline{(c) \hskip 5.8cm (d)}
 \caption{Simulation results with a real object: (a)~Object; 
Reconstruction from (b)~FT hologram, (c)~CFT hologram with $a_1 \approx 
a_2^\star$, and (d)~CFT hologram with $a_1=a_2^\star$.} \label{simul}
 \end{figure}

Although the derivations of the earlier sections are shown for 
continuous combined Fourier transform, it can be shown that the results 
are valid for discrete combined Fourier transform as well. We have 
carried out proof-of-the-principle study using computer simulations by 
using GNU Octave~\cite{gnu_octave}. The object 
transparency is of 200x200 pixels size as shown in Fig.\ref{simul}(a).
 The simulation window size is 256x256 pixels where the 200x200 pixels 
object is placed at the right hand top corner with zero padding. The 
object is purposefully off-centered so as to distinguish the twin 
reconstructions. The computational reconstruction of the object from a 
discrete FT hologram is carried out and is shown in Fig.\ref{simul}(b). 
Simulations for object reconstructions from the discrete CFT hologram
can be done for two cases: Case I considering real object, and Case II 
considering complex object.

\subsection{Case I: Reconstruction with a real object}

This case is straight forward. Computational reconstructions from 
simulated discrete CFT holograms for different values of $a_1$ and $a_2$ 
are shown in Fig.\ref{simul}(c) and \ref{simul}(d) for $a_1 \approx 
a_2^\star$ ($b_1=1.7$, $b_2=1$) and for $a_1=a_2^\star$ ($b_1=b_2=1$) 
respectively. It is evident from Fig.\ref{simul}(b) that reconstruction 
of the FT hologram reproduces twin image with equal strengths. On the 
otherhand, for the CFT hologram, the effect of twin image is present for 
$a_1 \neq a_2^\star$ ($b_1=1.7$, $b_2=1$), but the intensity levels of 
the two images are unequal [Fig.\ref{simul}(c)]. The effect of twin 
image is completely eliminated for $a_1=a_2^\star$ ($b_1=b_2=1$) as 
discernible from Fig.\ref{simul}(d). It is clear from the results of 
Fig.\ref{simul}(c) and Fig.\ref{simul}(d) that the effects of twin image 
in reconstructions of CFT holograms can be controlled by proper choice 
of $a_1$ and $a_2$, and can be completely eliminated for 
$a_1=a_2^\star$.

\subsection{Case II: Reconstruction with a complex object}

In practical situation the object would generally be complex, because 
the object transparency (either on glass plate or film) would have a 
phase variation due to variation in thickness or refractive index of the 
material or both. If $t(x,y)$ represents the thickness distribution over 
the object transparency, the phase distribution due to this thickness 
variation will be given by
 \begin{equation}
 \delta (x,y) = n \left({ {2 \pi}\over{\lambda} }\right) t(x,y)
 \end{equation}   

\noindent where $\lambda$ is the wavelength of the laser and $n$ is the 
refractive index of the material of the transparency. Here, it is 
assumed that the refractive index of the material of the object 
transparency is uniform. We consider an example of a quadratic thickness 
variation, so $t(x,y)$ may be expressed as
 \begin{equation}
 t(x,y) = t_0 + (x^2 + y^2) \Delta
 \end{equation}

\noindent where $t_0$ is the constant nominal thickness of the 
transparency plate and $\Delta$ is a thickness that is varied 
quadratically with spatial coordinates $x,y$. Therefore, the effect will 
be a phase factor $\exp[i\delta(x.y)]$ multiplied to the object 
function. If we put $\Delta=0$, we get a constant phase factor which is 
equivalent to the real object of Case I.

\begin{figure}[p] \hskip 1.5pt
 \centerline
 {\includegraphics[width=5.5cm]{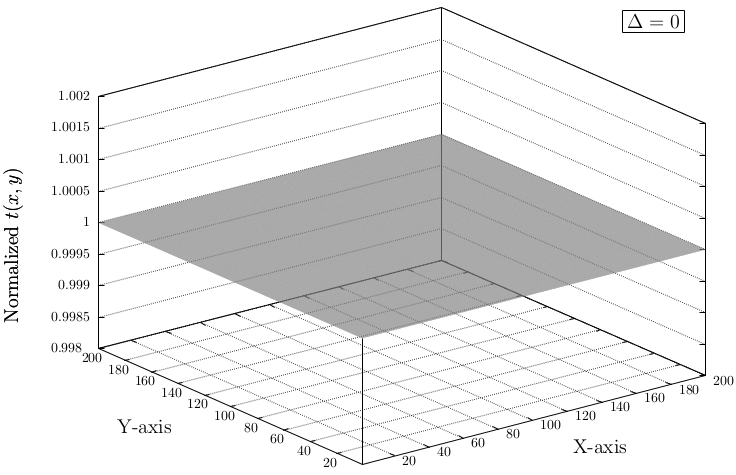} \hskip 24pt 
 \includegraphics[width=5.5cm]{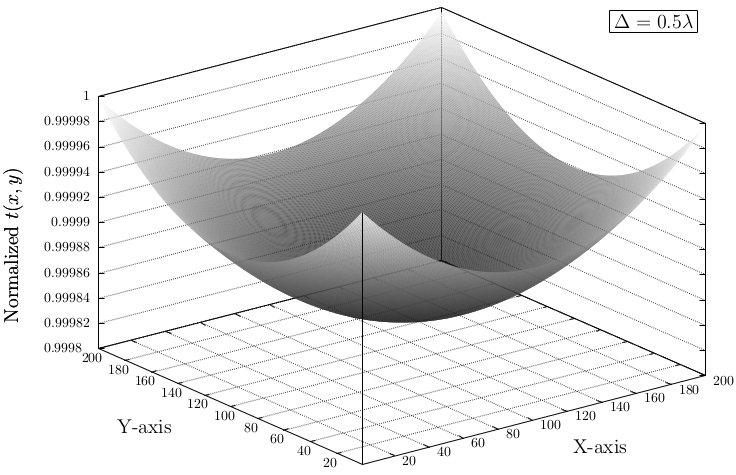} } \vskip 2pt
 \centerline{(a) \hskip 5.8cm (b)} \vskip 6pt 
 \centerline
 {\includegraphics[width=5.5cm]{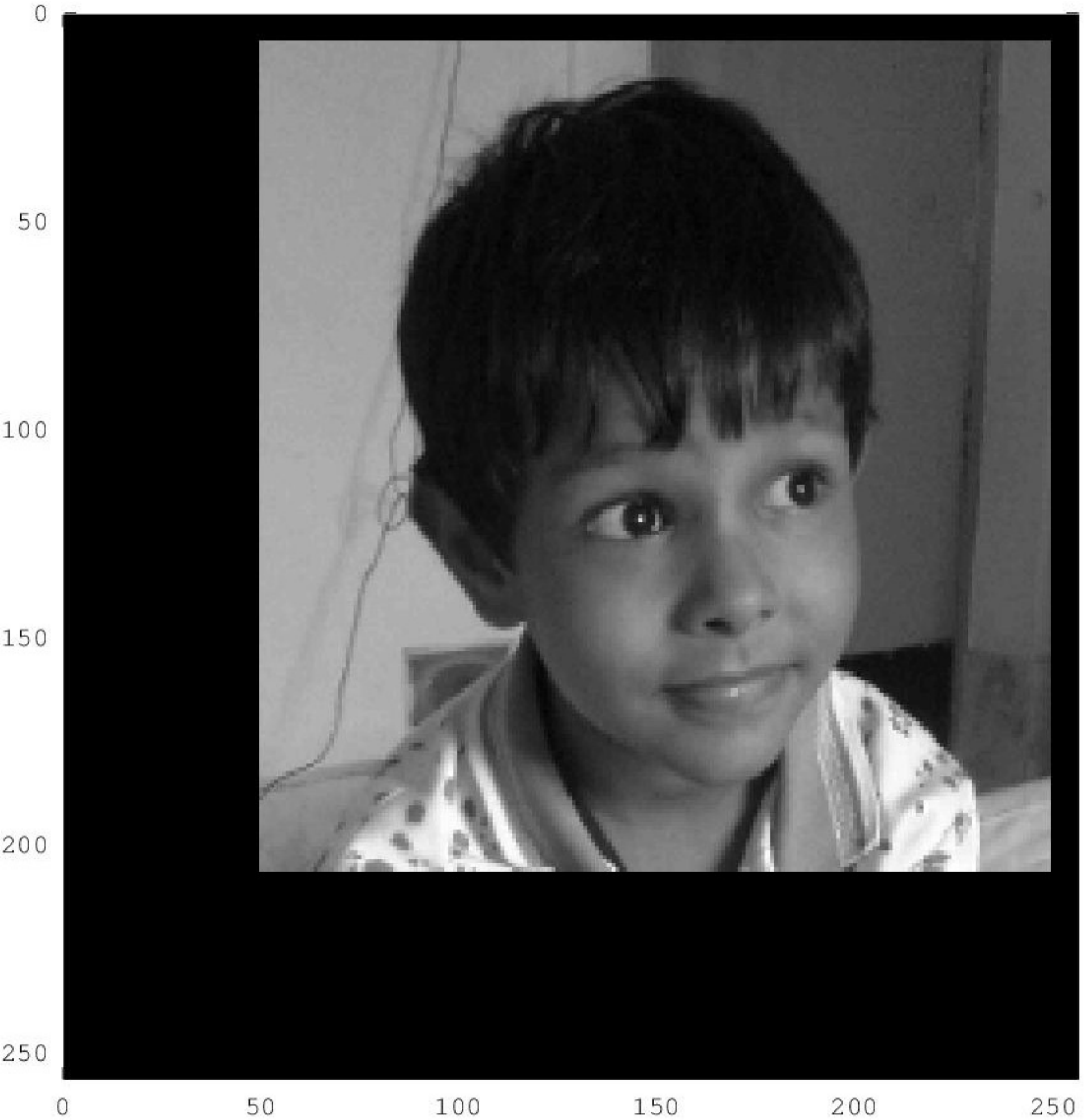} \hskip 24pt 
 \includegraphics[width=5.5cm]{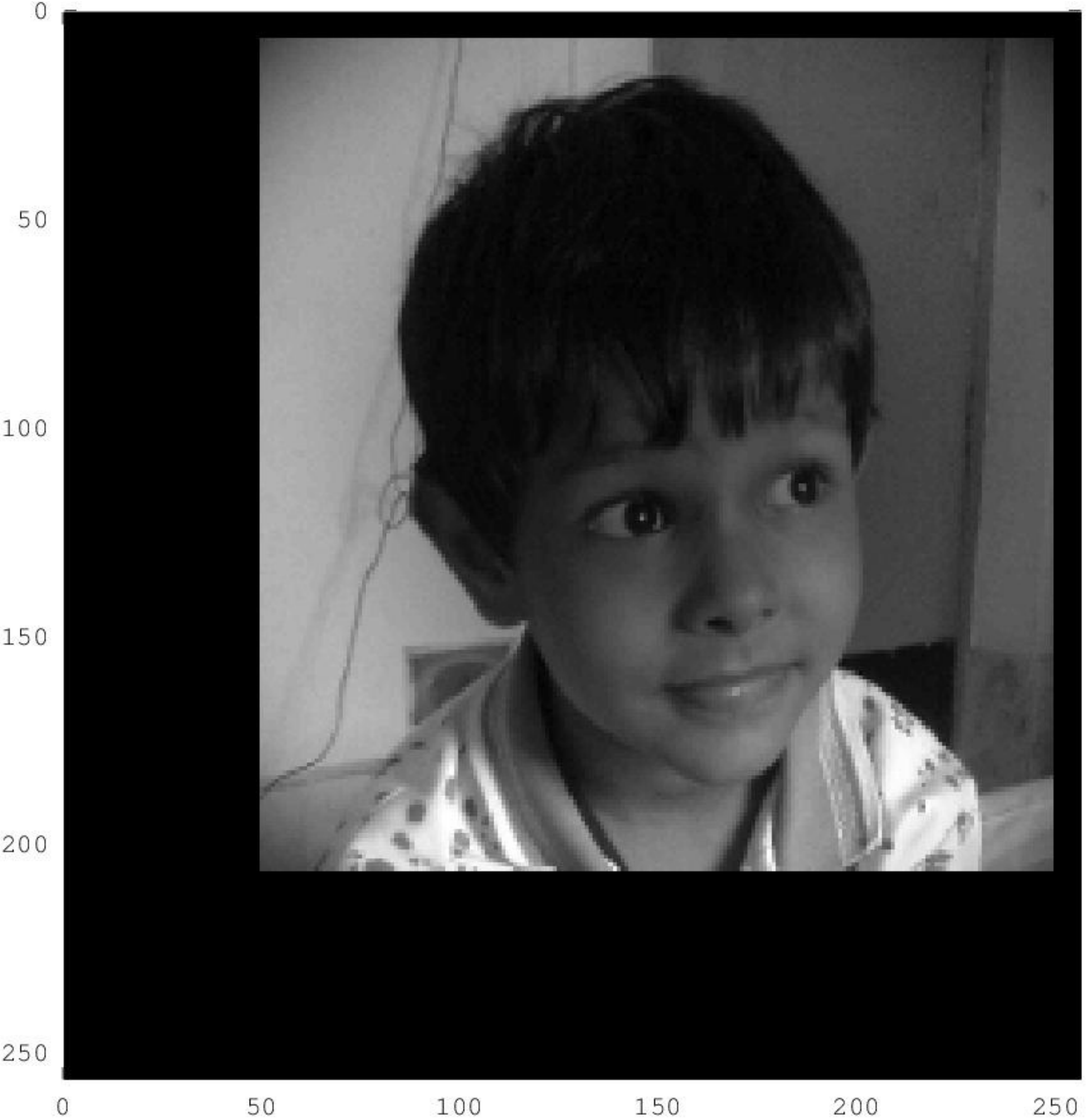} } \vskip 2pt
 \centerline{(c) \hskip 5.8cm (d)} \vskip 6pt
 \centerline
 {\includegraphics[width=5.5cm]{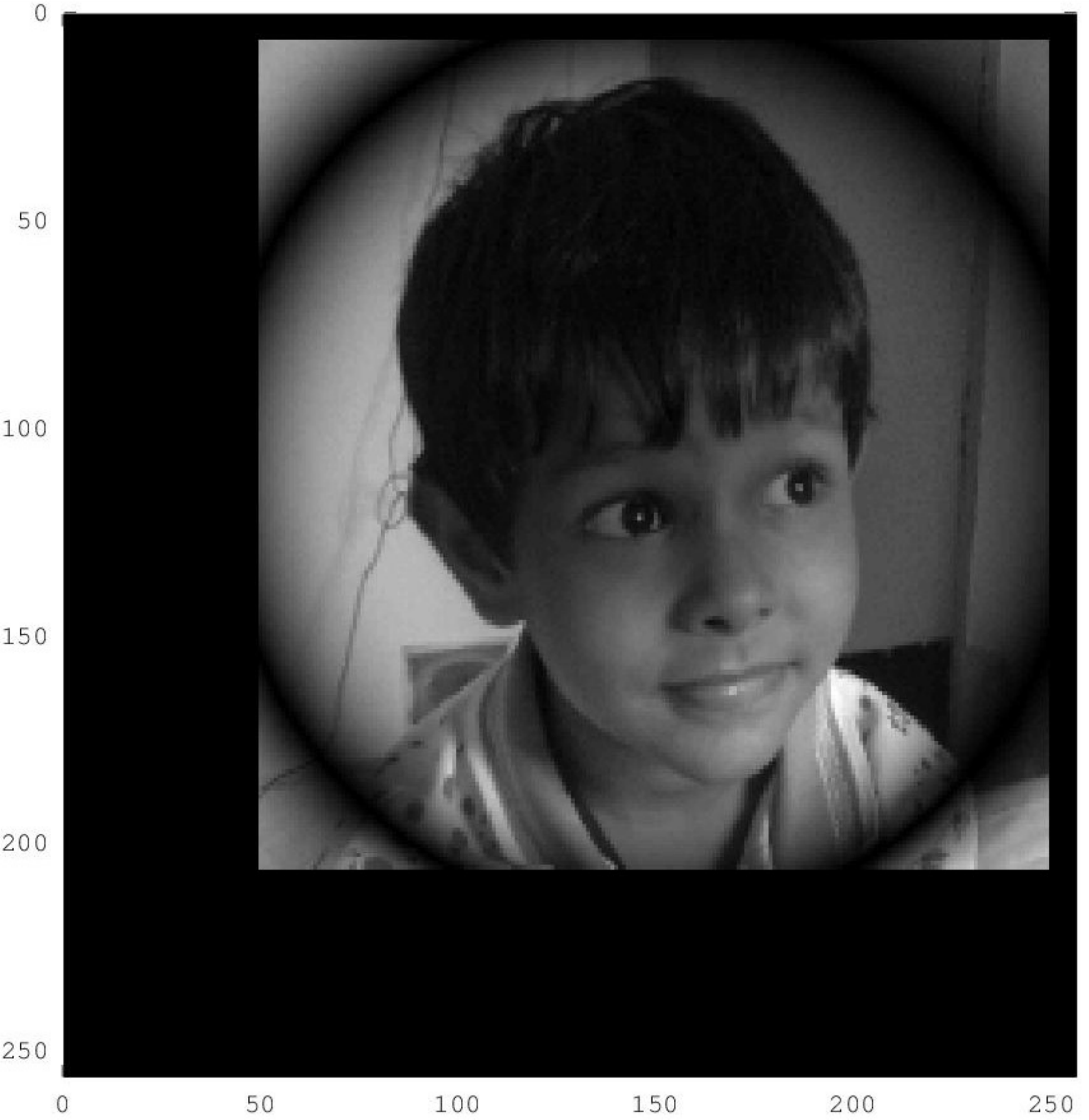} \hskip 24pt
 \includegraphics[width=5.5cm]{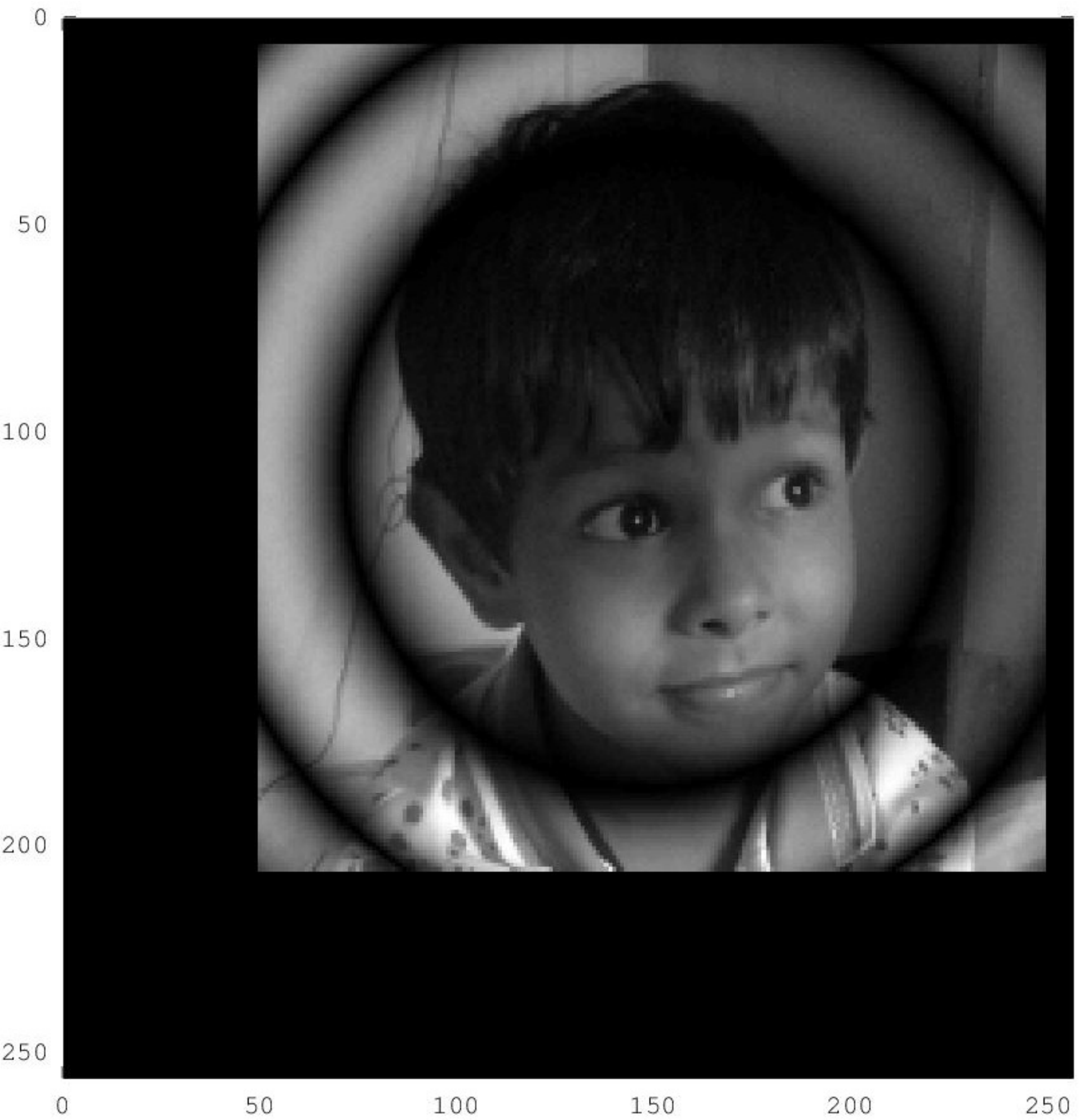} } \vskip 2pt
 \centerline{(e) \hskip 5.8cm (f)}
 \caption{Simulation results with a complex object: The normalized phase 
function $t(x,y)$ is shown for (a)~$\Delta=0$ and 
(b)~$\Delta=0.5\lambda$. Reconstructions from the CFT hologram are shown 
for (c)~$\Delta=0$, (d)~$\Delta=0.5\lambda$, (e)~$\Delta=\lambda$ and 
(f)~$\Delta=2\lambda$.} \label{simulc}
 \end{figure}

Simulations have been carried out with a 200x200 size phase function 
multiplied with the 200x200 size object function for $\lambda=633$~nm, 
$t_0=2$~mm and different values of $\Delta$, the overall simulation 
window remaining 256x256 size. The surface plot of the normalized phase 
function $t(x,y)$ for $\Delta=0$, and $\Delta=0.5\lambda$ are shown in 
Fig.\ref{simulc}(a) and \ref{simulc}(b). The phase function profile is 
constant in Fig.\ref{simulc}(a) which is nothing but the case of real 
object of Case I. The computational reconstructions of CFT holograms are 
shown in Fig.\ref{simulc}(c), \ref{simulc}(d), \ref{simulc}(e) and 
\ref{simulc}(f) for $\Delta=0$, 0.5$\lambda$, $\lambda$ and 2$\lambda$ 
respectively. Here, the weighting multipliers are kept at $a_1 = 
a_2^\star$ and $b_1 = b_2 = 1$ for all. It is evident from 
Fig.\ref{simulc}(c) -- \ref{simulc}(f) that the phase variation of the 
object transparency gives rise to circular fringes modulated over the 
reconstructed object image. The effect is negligible for 
$\Delta=0.5\lambda$ [Fig.\ref{simulc}(d)], but as $\Delta$ increases to 
$\lambda$, a circular fringe appears near the periphery of the image 
[Fig.\ref{simulc}(e)]. When $\Delta=2\lambda$, the image is modulated by 
two circular fringes [Fig.\ref{simulc}(f)]. It is clear that the 
thickness of the object transparency should be optically uniform, i.e., 
the surface finish of the object transparency plays an important role. 
That is why in such an interferometric arrangement, the object 
transparency may be placed inside an optical tank filled with an index 
matching liquid and optically polished glass windows with surface finish 
better than $\lambda/4$.

\section{Discussions}

The proof-of-the-principle simulations of the previous section proved 
the feasibility of twin image elimination in digital in-line holography 
using CFT. The proposed system is a polarization triangular path 
interferometer with a couple of Fourier lenses for synthesizing an 
inversion of the object function [$g(-x,-y)$], along with the object 
function [$g(x,y)$], and another Fourier lens for performing Fourier 
transform of the object and its inversion. The effect of CFT is created 
by introducing appropriate phase difference between the object 
information bearing waves by a polarization technique. Simulation is 
carried out for generating the CFT holograms for different values of the 
complex weighting factors. In this simulation, we have not considered 
the practical aspects of experimentation, such as effects of 
imperfections of the polarization components, aberrations of the Fourier 
lenses and the mirrors, misalignment errors of the interferometer, 
object wave to reference wave ratios and the quantization errors of the 
CCD camera.

The alignment of the proposed interferomeric system is not easy, 
nevertheless it is possible to implement combined Fourier transform 
optically. From equations~(\ref{a1a2:aphi}), (\ref{B}) and 
(\ref{cft:phi}) it is evident that the complex constants $a_1$ and $a_2$ 
are transformed into a phase difference $2\phi$. So, in optical 
implementation, the constants $a_1$ and $a_2$ are not required to be 
specified directly, instead the orientations of the polarizing 
components are so adjusted such that the phase difference $2\theta$ 
between the two image bearing object waves can be made equal to $2\phi$. 
If one desires, one can easily express the induced phase differences 
equivalent to the complex factors $a_1$ and $a_2$ using De Moivre's 
relations of equations~(\ref{a1a2:aphi}) and (\ref{B}). The complex 
coefficients $a_1$ and $a_2$ are also identified in the output of the 
interferometer in equation~(\ref{pi}) as exponential phase factors.

Since, it is an interferometric transformation, the holographic 
reconstruction is sensitive to phase aberrations of the object 
transparency. Thus, this technique may be useful for carrying out 
studies of phase objects. A similar inverting triangular path 
polarization interferometer was indeed utilized for implementing optical 
Hartley transform experimentally~\cite{debHT:jopt1997}. We wish to 
consider the practical aspects in a future experimental implemetation of 
the proposed method.

\section{Conclusion}
 We have proposed a technique for resolving twin image problem of 
in-line digital Fourier holography. We use a combination of Fourier 
transforms to eliminate the unwanted twin image. The technique relies on 
interferometric addition of Fourier transforms of the object and its 
inversion with appropriate complex weighting coefficients, which are 
introduced through polarization-induced phase. Simulation results prove 
the feasibility of complete elimination of one of the twin image. 
Moreover, the technique is sensitive to phase aberrations of the object 
transparency, thereby providing a way for phase sensing holographic 
imaging. The proposed method neither involves recording of several phase 
shifted holograms, nor it requires any iteration for digital 
reconstruction, which may render it suitable for recording holograms of 
fast changing sequences as well as fast digital reconstruction of the 
recorded holograms.

 \begin{acknowledgements}
 The authors thank an anonymous reviewer for fruitful comments that 
helped to improve the paper.
 \end{acknowledgements}


\begin{thebibliography}{99}

\bibitem{gabor} D. Gabor, ``A new microscopic principle,'' Nature {\bf 
161}, 777-778 (1948).

\bibitem{rogers:nature51} W. L. Bragg and G. L. Rogers, ``Elimination of 
the unwanted image in diffraction microscopy,'' Nature {\bf 167}, 190 
(1951).

\bibitem{twin:rev} B. M. Hennelly, D. P. Kelly, N. Pandey, D. Monaghan, 
``Review of twin reduction and twin removal techniques in holography,'' 
Proc. China-Ireland International Conference on Information and 
Communications Technologies, Maynooth, Ireland, Pages 241-245 (2009).

\bibitem{leith_upatnieks} E. Leith and J. Upatnieks, ``Wavefront 
reconstruction with continuous-tone objects,'' JOSA {\bf 53} 1377-1381 
(1963).

\bibitem{cft:ansari} R. Ansari, ``An extension of the discrete Fourier 
transform,'' IEEE Trans Circuits Sys. {\bf CAS-32} (6), 618-619 (1985).

\bibitem{deb:polaphase} D. Choudhury, P. N. Puntambekar and A. K. 
Chakraborty, ``Utilization of polarization-induced phase difference for 
complex addition and subtraction of amplitudes,'' J. Opt. {\bf 22}, 6-10 
(1993).

\bibitem{jones:calculus} A. Gerald and J.M. Burch, {\em Introduction to 
Matrix Methods in Optics} (John Wiley \& Sons, 1975).

\bibitem{gnu_octave} GNU Octave: http://www.gnu.org/software/octave/ 
(accessed December 4, 2012)

\bibitem{debHT:jopt1997} D. Choudhury, P. N. Puntambekar and A. K. 
Chakraborty, ``Hartley transformation by an inverting interferometer,'' 
J. Opt. {\bf 26} (1997) 139-145.


\end{thebibliography}
\end{document}